\documentclass[12pt]{iopart}
\usepackage{graphicx}
\begin{document}
\title[Systematic study of isoscaling behavior in projectile fragmentation]
{Systematic study of isoscaling behavior in projectile fragmentation
by the statistical abrasion-ablation model}

\author{D. Q. Fang\footnote{Corresponding author.
{\it Email address: dqfang@sinap.ac.cn}}
, Y. G. Ma, C. Zhong, C. W. Ma,
X. Z. Cai, J. G. Chen, W. Guo, Q. M. Su, W. D. Tian, K. Wang,
T. Z. Yan, W. Q. Shen
}

\address{Shanghai Institute of Applied Physics, Chinese Academy
of Sciences, Shanghai 201800, People's Republic of China}

\begin{abstract}
The isospin effect and isoscaling behavior in projectile fragmentation have
been systematically investigated by a modified statistical abrasion-ablation
(SAA) model. The normalized peak differences and reduced isoscaling parameters
are found to decrease with ($Z_{\mbox{proj}}-Z$)/$Z_{\mbox{proj}}$ or the
excitation energy per nucleon and have no significant dependence on the size
of reaction systems. Assuming a Fermi-gas behavior, the excitation energy
dependence of the symmetry energy coefficients are tentatively extracted
from $\alpha$ and $\beta$ which looks consistent with the experimental data.
It is pointed out that the reduced isoscaling parameters can be used as an
observable to study excitation extent of system and asymmetric nuclear
equation of state in heavy ion collisions.
\end{abstract}



\section{Introduction}

The process of projectile fragmentation has been studied extensively
for investigation of reaction mechanisms in heavy ion collisions
at intermediate and high energies \cite{BOW,HUF,GOS,MOR,DAY,BON,Brohm1994NPA}.
It is also one of the most important methods to produce exotic nuclei.
Recent advances in experiments using radioactive ion beams with
large neutron or proton excess have lead to the discovery of
halo structure \cite{TAN1,TAN2}. Since then interest in the study of
very neutron-rich and proton-rich nuclei has grown due to their
anomalous structures. In addition, the studies of isospin physics have become
a very popular subject. The isospin effects of various
physical phenomena, such as multifragmentation, flow, pre-equilibrium
nucleon emission, etc., have been extensively reported
\cite{MIL,DEM,LIB1,LIB2,PAK,KUM,Muller1995,Ma1999}.
The studies have shown that isospin effect exists in nuclear reactions
induced by exotic nuclei but it may disappear under certain conditions.
Our previous calculations by using the modified statistical abrasion-ablation
(SAA) model have demonstrated that the fragment isotopic distribution shifts toward
the neutron-rich side for neutron-rich projectile, but the shift decreases
with the increase of the parameter (Z$_{\mbox{proj}}$-Z)/Z$_{\mbox{proj}}$
or the violence of nuclear reaction. This isospin effect of
fragmentation reaction on the fragment isotopic distribution will disappear
when (Z$_{\mbox{proj}}$-Z)/Z$_{\mbox{proj}}$ becomes larger than 0.5
\cite{Fang2001CPL,Fang2000PRC}.

Recently, study of the nuclear symmetry energy has become a very important
topic in nuclear physics. It is well known that the nuclear symmetry energy
is very significant for investigation of the nuclear equation of state and
a variety of astrophysical phenomena.
The isoscaling approach for light fragment composition produced in the
multifragmentation of very hot source has become an important method
in heavy ion collisions since it can isolate the nuclear symmetry energy
in the fragment yields \cite{Tsang2001PRL,Tsang2001PRC1,Tsang2001PRC2,Botvina}.
The scaling law relates ratios of isotope yields measured in two
different nuclear reactions, 1 and 2,
$R_{21}(N,Z)=Y_2(N,Z)/Y_1(N,Z)$. In multifragmentation events,
such ratios are shown to obey an exponential dependence on the
neutron number $N$ or proton number $Z$ of the isotopes or isotones
characterized by three parameters $\alpha$, $\beta$ and C
\cite{Tsang2001PRL}:
\begin{equation}
   R_{21}(N,Z) = \frac{Y_2(N,Z)}{Y_1(N,Z)} = C \exp(\alpha N + \beta
    Z),
\end{equation}
here $C$ is an overall normalization constant.
In the grand-canonical limit, $\alpha$ and $\beta$ will have the form,
\begin{equation}
\alpha =
\frac{4C_{\mbox{sym}}}{T}[(\frac{Z_1}{A_1})^2-(\frac{Z_2}{A_2})^2]
\equiv
\frac{4C_{\mbox{sym}}}{T}\Delta[(\frac{Z}{A})^2]
\end{equation}
and
\begin{equation}
\beta =
\frac{4C_{\mbox{sym}}}{T}[(\frac{N_1}{A_1})^2-(\frac{N_2}{A_2})^2]
\equiv
\frac{4C_{\mbox{sym}}}{T}\Delta[(\frac{N}{A})^2],
\end{equation}
where $C_{\mbox{sym}}$ is symmetry energy coefficient (MeV),
$(\frac{Z_{\mbox{i}}}{A_{\mbox{i}}})^2$ or
$(\frac{N_{\mbox{i}}}{A_{\mbox{i}}})^2$ (i=1,2)
means the square of charge or neutron number over mass number for system 1 and 2.
$T$ is the temperature of the system in MeV. This behavior is
attributed to the difference of isospin asymmetry between two
reaction systems in similar nuclear temperature.
Since the symmetry energy determines nuclear structure of neutron-rich
or neutron-deficient rare isotopes,
studies on the isoscaling behavior can be used
to probe the isospin dependent nuclear equation of
state \cite{Tsang2001PRL,Tsang2001PRC1,Tsang2001PRC2,Botvina,DiToro,
Ma_review,Ma2004PRC,Tian,FevrePRL,Souliotis05}.

So far, the isoscaling behavior has been studied  experimentally
and theoretically for different reaction mechanisms. However, most
studies focus on the isoscaling behaviors for light particles with
$Z$=2$-$8. A few studies on the heavy projectile-like residues
in deep elastic collisions and fission fragments have been reported
\cite{Souliotis05,Souliotis2003PRC,Friedman,Veselsky2,Wang,Ma_fis}.

In this paper, we will present our studies on systematic
behaviors of the isospin effect as well as isoscaling features for
projectile-like fragments in the framework of statistical
abrasion-ablation model. Extraction of the symmetry energy
coefficient from the isoscaling parameters will also be investigated.

\section{Model description}

The statistical abrasion-ablation  model can describe the
isotopic distribution well \cite{Brohm1994NPA}. In the SAA model ,
the nuclear reaction is described as two stages which occur in two
distinctly different time scales. The first abrasion stage is
fragmentation reaction which describes the production of the
pre-fragment with certain amount excitation energy through the
independent nucleon-nucleon collisions in the overlap zone of the
colliding nuclei.  The collisions are described by a picture of
interacting tubes. Assuming a binomial distribution for the
absorbed projectile neutrons and protons in the interaction of a
specific pair of tubes, the distributions of the total abraded
neutrons and protons are determined. For an infinitesimal tube in
the projectile, the transmission probabilities for neutrons
(protons) at a given impact parameter $b$ are calculated by
\cite{Brohm1994NPA}
\begin{equation}
t_{\mbox{k}}(\vec{r}-\vec{b})=\exp\{-[D_{\mbox{n}}^{\mbox{T}}
(\vec{r}-\vec{b})\sigma_{\mbox{nk}}+%
D_{\mbox{p}}^{\mbox{T}}(\vec{r}-\vec{b})\sigma_{\mbox{pk}}]\},
\end{equation}
where $D^{\mbox{T}}$ is thickness function of
the target, which is normalized by
$\int d^2rD^{\mbox{T}}_{\mbox{n}}=N^{\mbox{T}}$ and
$\int d^2rD^{\mbox{T}}_{\mbox{p}}=Z^{\mbox{T}}$ with
$N^{\mbox{T}}$ and $Z^{\mbox{T}}$ referring to the neutron and
proton number in the target respectively, the vectors $\vec{r}$ and $\vec{b}$
are defined in the plane perpendicular to beam, and
$\sigma_{\mbox{k}'\mbox{k}}$ is the free nucleon-nucleon cross
sections (k$'$, k$=$n for neutron and k$'$, k$=$p for proton). The
thickness function of the target is given by
\begin{equation}
D^{\mbox{T}}_{\mbox{k}}(r)=\int_{-\infty}^{+\infty}dz\rho_{\mbox{k}}
((r^2+z^2)^{1/2}),
\end{equation}
with $\rho_{\mbox{k}}$ being the neutron (proton) density
distribution of the target. So the average abraded mass at a given
impact parameter $b$ is calculated by the expression
\begin{equation}
\begin{array}{ll}
\langle \Delta A(b) \rangle= & \int d^2rD_{\mbox{n}}^{\mbox{P}}(r)
[1-t_{\mbox{n}}(\vec{r}-\vec{b})] \\
&+\int d^2rD_{\mbox{p}}^{\mbox{P}}(r)[1-t_{\mbox{p}}(\vec{r}-\vec{b})].
\end{array}
\end{equation}

\begin{figure}[t]
\begin{center}
\includegraphics[width=6.2cm,angle=-90]{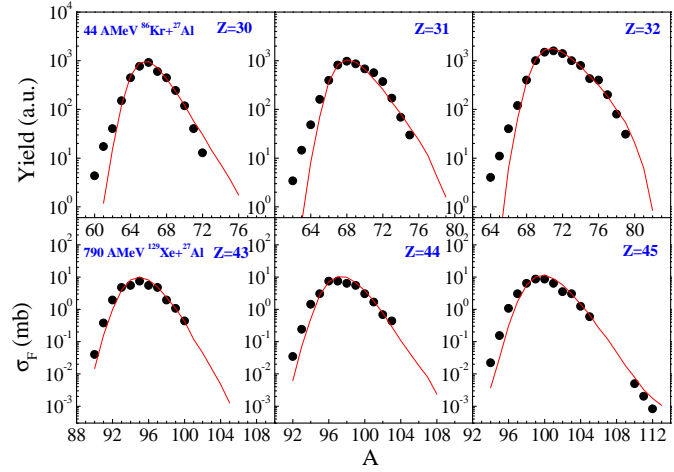}
\caption{Comparison of isotopic distributions between the SAA model and the data.
The isotopic distributions for selected charge numbers from
44 $A$MeV $^{86}$Kr+$^{27}$Al (upper panel) and
790 $A$MeV $^{129}$Xe+$^{27}$Al (lower panel). The dots are the experimental data
taken from Ref.\cite{BAZ} for 44 $A$MeV $^{86}$Kr+$^{27}$Al
and Ref.\cite{REI} for 790 $A$MeV $^{129}$Xe+$^{27}$Al,
the lines are the results calculated by the SAA model.}
\label{exp}
\end{center}
\end{figure}

\begin{figure}[t]
\begin{center}
\includegraphics[width=8cm]{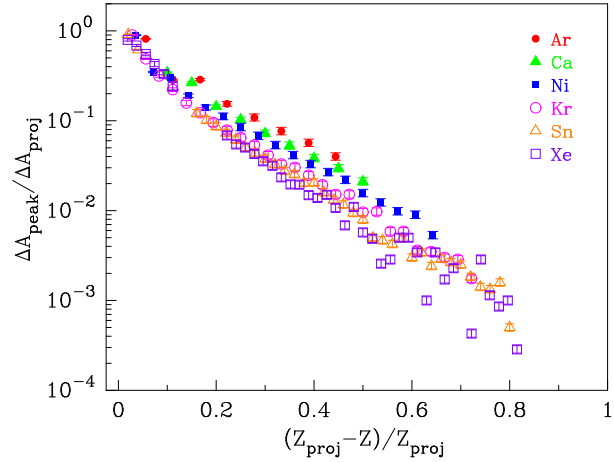}
\caption{The normalized peak difference
$\Delta$A$_{\mbox{peak}}$/$\Delta$A$_{\mbox{proj}}$
of the fragment isotopic distribution as a function of
($Z_{\mbox{proj}}-Z$)/$Z_{\mbox{proj}}$. Different
symbols are used for projectiles with different charge number
as shown in the legend.}
\label{iso}
\end{center}
\end{figure}

\begin{figure}[t]
\vspace{0.2cm}
\begin{center}
\includegraphics[width=15.5cm]{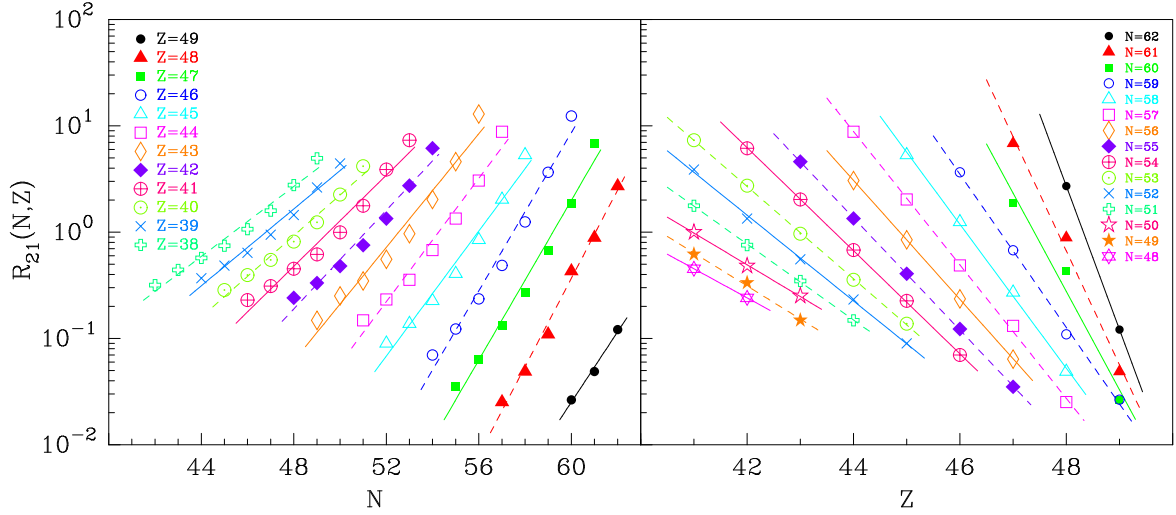}
\caption{Yield ratios $R_{21}(N,Z)$ of fragments from
the reactions of $^{124/112}$Sn+$^{112}$Sn at 60$A$ MeV versus  $N$ for
the selected isotopes (left panel) and $Z$ for the selected isotones
(right panel). Different symbols are used for different isotopes and
isotones as shown in the legend. The lines represent the exponential fits.
For details see text.}
\label{R21_Proton_Neutron}
\end{center}
\end{figure}

The second stage is the evaporation process in which the system
reorganizes due to excitation. It deexcites and
thermalizes by the cascade evaporation of light particles using
the conventional statistical model~\cite{Gimmard}.
The excitation energy for projectile spectator is estimated by a
simple relation of $E^* = 13.3 \langle \Delta A(b) \rangle$ MeV
where 13.3 is a mean excitation energy for an abraded nucleon
from the initial projectile. This excitation energy was given
by the statistical hole-energy model as described in Ref.~\cite{Gimmard}.
After the evaporation stage, we can obtain the final fragments
which are comparable to the experimental data.
By introducing in-medium nucleon-nucleon cross section and optimizing
computational method given in
Ref. \cite{Cai2002PRC,Fang2001CPL,Fang2000PRC,Zhong2003HEP},
it can give a good agreement with the experimental isotopic distributions
\cite{Fang2001CPL,Fang2000PRC,Zhong2003HEP}. Comparison of the SAA model
calculations with the experimental isotopic distributions for
$Z=30-32$ from  44 $A$MeV $^{86}$Kr+$^{27}$Al \cite{BAZ} and
$Z=43-45$ from 790 $A$MeV $^{129}$Xe+$^{27}$Al \cite{REI} is shown
in Fig.~\ref{exp}.
The results shown in this figure and all the following figures are
referring the final fragments after the evaporation stage.
For $^{86}$Kr, all isotopic distributions are normalized
by the same factor in order to compare with the experimental yields.
For $^{129}$Xe, the calculated production cross sections
are compared with the data directly. From this figure, we can see that
the SAA model can reproduce the experimental data both at intermediate
and high energies quite well.
The isospin effect and its disappearance in projectile fragmentation
for $^{36,40}Ar$ at intermediate energies have been predicted by
this model and confirmed by the experimental data \cite{Fang2000PRC}.

\section{Calculations and discussion}

The model predicts that strong isospin effect exists in the isotopic
distributions produced by projectiles with same charge number but different
mass number \cite{Fang2001CPL,Fang2000PRC}.
In order to do a systematic study of the isospin effect
in projectile fragmentation, reactions of $^{40/36}$Ar,  $^{48/40}$Ca,
$^{64/58}$Ni, $^{86/78}$Kr, $^{124/112}$Sn and $^{129/136}$Xe on $^{112}$Sn
at 60$A$ MeV are simulated by the SAA model.
Since the isotopic distributions from two projectiles have similar shape but a shift
in mass, their peak positions will be one of the most sensitive quantities for
the isospin effect. Thus we extract the peak position by
Gaussian fit to the fragment isotopic distribution for each charge number Z
as in Ref.\cite{Fang2001CPL,Fang2000PRC}.
The normalized differences of the peak position from two projectiles
$\Delta$A$_{\mbox{peak}}$/$\Delta$A$_{\mbox{proj}}$
as a function of 
($Z_{\mbox{proj}}-Z$)/$Z_{\mbox{proj}}$ are shown in Fig.~\ref{iso}.
Here $\Delta$A$_{\mbox{proj}}$ is the mass number difference between
the two projectiles with same charge number and $\Delta$A$_{\mbox{peak}}$ is the peak
position difference of the fragment isotopic distribution produced by
these two projectiles.
$Z_{\mbox{proj}}$ and $Z$ are the charge number of the projectile and the
produced isotopes.
$\Delta$A$_{\mbox{peak}}$/$\Delta$A$_{\mbox{proj}}$
exponentially decreases as the increase of ($Z_{\mbox{proj}}-Z$)/$Z_{\mbox{proj}}$
which is same as our previous conclusions \cite{Fang2000PRC}. The dependence of
$\Delta$A$_{\mbox{peak}}$/$\Delta$A$_{\mbox{proj}}$ on
($Z_{\mbox{proj}}-Z$)/$Z_{\mbox{proj}}$
shows a very slight difference among different size projectiles.

To study systematic behaviors of the isoscaling phenomena, the yield
ratios $R_{21}(N,Z)$ are made using the convention that index 2 refers to
the more neutron-rich system and index 1 to the less neutron-rich one.
As an example, Fig.~\ref{R21_Proton_Neutron} shows
the yield ratios $R_{21}(N,Z)$
as a function of neutron number $N$ for selected isotopes and $Z$ for
selected isotones from $^{124/112}$Sn + $^{112}$Sn reactions in log-scale.
From this figure, we observe that the ratio for
each isotope $Z$ exhibits a remarkable exponential behavior.
For each isotope ($Z$), an exponential function form $C \exp(\alpha N)$ is
used to fit the calculated points and the parameters $\alpha$ are obtained
for all isotopes. Analogous behavior is observed for each isotone ($N$),
an exponential function form $C' \exp(\beta Z)$ is used to fit the
calculated points and the parameters $\beta$ are obtained for all isotones.

\begin{figure}[h]
\begin{center}
\includegraphics[width=8cm]{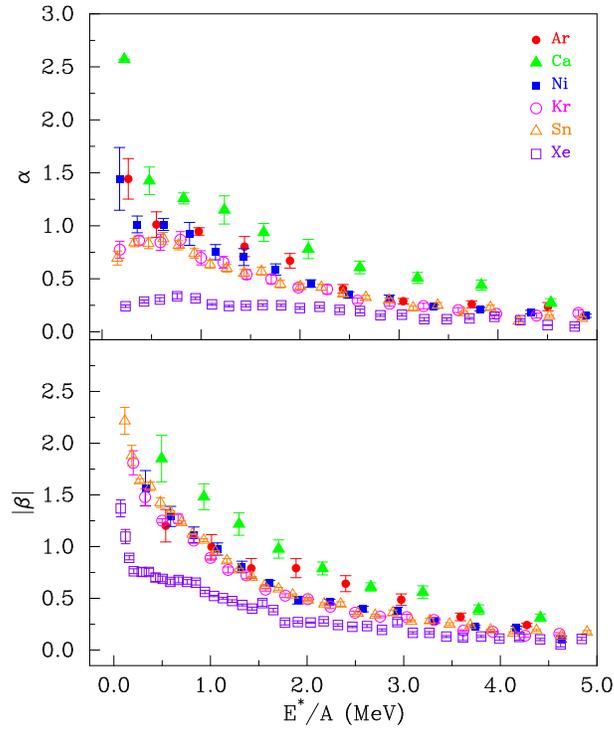}
\caption{Same as in Fig.~\ref{iso} but for the isoscaling
parameters $\alpha$ (upper panel) and $|\beta|$ (lower panel) as a
function of the excitation energy per nucleon.}
\label{alpha_beta}
\end{center}
\end{figure}

\begin{figure}[t]
\begin{center}
\includegraphics[width=8cm]{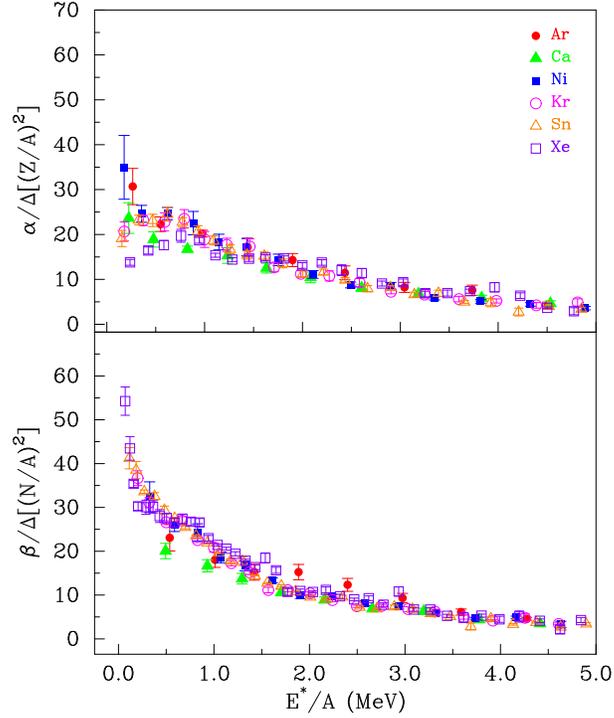}
\caption{Same as in Fig.~\ref{iso} but for the reduced isoscaling
parameters $\alpha$/$\Delta[(\frac{Z}{A})^2]$ (upper panel) and
$\beta$/$\Delta[(\frac{N}{A})^2]$ (lower panel) as a
function of the excitation energy per nucleon.}
\label{sub}
\end{center}
\end{figure}

In Fig.~\ref{alpha_beta}, we present the extracted slope parameters
$\alpha$ (upper panel) and $|\beta|$ (lower panel) of the exponential
fits as a function of the excitation energy per nucleon ($E^*/A$).
Since the excitation energy changes as a function of time in the
process of evaporation, the values of $E^*/A$ shown here are
taken at the beginning of the evaporation stage and
the mass of the prefragment is used to calculated $E^*/A$.
In the model the excitation energy is proportional
to the abraded nucleons and can reflect the violence of the collision as
the parameter ($Z_{\mbox{proj}}-Z$)/$Z_{\mbox{proj}}$ shows \cite{Fang2000PRC}.
In this figure, $\alpha$ and $|\beta|$ show a decreasing trend with the
increasing of $E^*$/$A$.
This behavior for projectile-like fragments is different with light particles.
The isoscaling parameters of light fragments from multifragmentation is almost
constant for different isotopes because the excitation energy or temperature
is almost same for all light fragments in the process of multifragmentation.
The decrease of the isoscaling parameters in our calculations
may mainly be attributed to the evaporation effect of the prefragment
with different excitation energy as in the disappearance
of the isospin effect \cite{Fang2001CPL,Fang2000PRC}.
The values of $\alpha$ and $|\beta|$ are quite different from
different reaction systems due to the different size and isospin of
the projectiles.

According to Eq.~(2) and (3), $\alpha$ and $\beta$ have a linear dependence on
$\Delta[(\frac{Z}{A})^2]$ or $\Delta[(\frac{N}{A})^2]$.
Since this parameter is dependent on the reaction system,
we divide $\alpha$ ($\beta$) by $\Delta[(\frac{Z}{A})^2]$
($\Delta[(\frac{N}{A})^2]$) to remove the system isospin and size
dependence and call them reduced isoscaling parameters.
The results are given in Fig.~\ref{sub}. After the reduction,
$\alpha$/$\Delta[(\frac{Z}{A})^2]$ ($\beta$/$\Delta[(\frac{N}{A})^2]$)
of different reaction systems demonstrate almost same dependence with $E^*$/$A$.
Eq.~(2) and (3) are deduced from the grand-canonical limit for
multifragmentation of hot source. For projectile-like fragments,
the same behavior is observed in the SAA model. In this sense,
the reduced isoscaling parameters
$\alpha$/$\Delta[(\frac{Z}{A})^2]$ ($\beta$/$\Delta[(\frac{N}{A})^2]$)
may be used as a sensitive observable for measuring the excitation extent of
projectile-like fragments during the collisions without system size dependence.

\begin{figure}[t]
\begin{center}
\includegraphics[width=8cm]{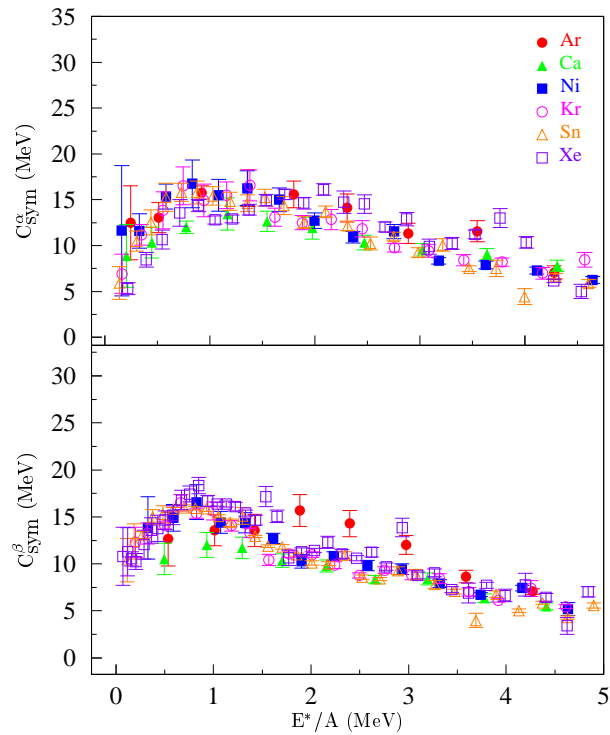}
\caption{Same as in Fig.~\ref{iso} but for the extracted symmetry
energy coefficient from $\alpha$ ($C^{\alpha}_{\mbox{sym}}$, upper panel) and
$\beta$ ($C^{\beta}_{\mbox{sym}}$, lower panel) as a function of the
excitation energy per nucleon.}
\label{Csym}
\end{center}
\end{figure}

From above discussions, we found that
$\Delta$A$_{\mbox{peak}}$/$\Delta$A$_{\mbox{proj}}$ and
$\alpha$/$\Delta[(\frac{Z}{A})^2]$ ($\beta$/$\Delta[(\frac{N}{A})^2]$)
decrease with ($Z_{\mbox{proj}}-Z$)/$Z_{\mbox{proj}}$ or
the excitation energy per nucleon.
But the later one decreases slower than the first one. It means
that $\alpha$/$\Delta[(\frac{Z}{A})^2]$ ($\beta$/$\Delta[(\frac{N}{A})^2]$)
is more sensitive to the isospin effect of the projectiles.
Since $\alpha$/$\Delta[(\frac{Z}{A})^2]$ ($\beta$/$\Delta[(\frac{N}{A})^2]$)
is related to $\frac{C_{\mbox{sym}}}{T}$ as in Eq.~(2) and (3),
it can be used as an observable to study the excitation and asymmetric
nuclear equation of state in heavy ion collisions.

If we use the Fermi-gas relationship between excitation
energy per nucleon and temperature $E^*/A=\frac{1}{a}T^2$
to calculate $T$ tentatively, with the inverse level density parameter
$a$=8$-$13 (in our calculation $a$=10 is used),
the symmetry energy coefficient ($C_{\mbox{sym}}$) could be extracted.
Results extracted from
$\alpha$/$\Delta[(\frac{Z}{A})^2]$ and $\beta$/$\Delta[(\frac{N}{A})^2]$
are shown in Fig.~\ref{Csym}. For $E^*/A$ around
1 MeV, the symmetry energy coefficients from $\alpha$ and $\beta$ are around 15 MeV.
These values are a little lower than the standard value
$C_{\mbox{sym}}$=25 MeV in liquid drop model \cite{Botvina},
but it seems consistent with the extracted results from the experimental
data by A. Le F\'{e}vre {\it et al.} \cite{FevrePRL}.
The obtained $C_{\mbox{sym}}$ is not a constant and decreases with the
increase of $E^*/A$. Similar dependence was also observed in the experimental
studies but their $C_{\mbox{sym}}$ values is a little bit larger than
ours \cite{FevrePRL,Souliotis05}.
It should be pointed out that the experimental data are taken at different
incident energies (300 $A$ MeV and 600 $A$ MeV in Ref. \cite{FevrePRL}, around 25 $A$ MeV in
Ref. \cite{Souliotis05}). Our calculations are performed at 60 $A$ MeV, but we have
found that there is almost no incident energy dependence for the isoscaling parameters
in our model.
In our results, the symmetry energy coefficient decreases quickly
when $E^*/A$ is less than 1 MeV. This may stem from the increase of the
inverse level density parameter at low excitation energy \cite{NATOWIZPRC02}.
Of course, in this low $E^*/A$ range there is very few theoretical and
experimental data up to now and more researches are necessary.
Experimentally it may be difficult to extract $E^*/A$. As we have mentioned
previously that the parameter ($Z_{\mbox{proj}}-Z$)/$Z_{\mbox{proj}}$
can reflect the violence of nuclear collision and is approximately
proportional to $E^*/A$ in not very central collisions.
Actually quite similar dependence as in
Fig.~\ref{alpha_beta}-\ref{Csym} is seem if
($Z_{\mbox{proj}}-Z$)/$Z_{\mbox{proj}}$ is used instead of $E^*/A$.
Thus we can also study the dependence of the isoscaling parameter
and symmetry energy coefficient with ($Z_{\mbox{proj}}-Z$)/$Z_{\mbox{proj}}$
experimentally when there is no $E^*/A$ data.

However, some cautions should be reminded.
In the SAA model, the symmetry energy term is not taken into account explicitly.
But a similar analysis of isoscaling as in the
statistical multifragmentation model could be done since
there exists different isotopic and isotonic distributions
between two systems.
Of course, the effect of symmetry energy term should have been
reflected implicitly in the assumption of the abrasion and also
evaporation stages in the SAA model. Thus a clear isoscaling
behavior is observed for projectile-like fragments in the
this work and the symmetry energy coefficient extracted
based on Eq. (2) and (3) is consistent with the experimental
data~\cite{FevrePRL,Souliotis05}.
The present calculation could provide some useful information for
further experimental and theoretical investigations on the isoscaling
of projectile-like fragments.

\section{Summary}
In summary, systematic behaviors of the isospin effect and isoscaling
of projectile-like fragments from $^{40/36}$Ar,  $^{48/40}$Ca,
$^{64/58}$Ni, $^{86/78}$Kr, $^{124/112}$Sn and $^{129/136}$Xe on $^{112}$Sn
at 60$A$ MeV have been studied by a modified statistical abrasion-ablation
model. The normalized peak differences
$\Delta$A$_{\mbox{peak}}$/$\Delta$A$_{\mbox{proj}}$
for different reaction systems show similar dependence with the parameter
($Z_{\mbox{proj}}-Z$)/$Z_{\mbox{proj}}$.
The isoscaling parameters $\alpha$ and $\beta$ are extracted for
the produced isotopes and isotones, and they show different values for
different systems. However, the reduced isoscaling parameters
$\alpha$/$\Delta[(\frac{Z}{A})^2]$ and $\beta$/$\Delta[(\frac{N}{A})^2]$
show almost same dependence with $E^*/A$ for different systems.
Assuming a Fermi-gas behavior, the symmetry energy coefficients are tentatively
extracted from $\alpha$ and $\beta$ and it seems that the results are consistent
with the experimental data.

\ack
This work was partially supported by the National Natural Science
Foundation of China (NNSFC) under Grant No. 10405032, 10535010,
10405033 and 10475108, Shanghai Development
Foundation for Science and Technology under contract No. 06QA14062,
06JC14082 and 05XD14021, the Major State Basic
Research Development Program in China under Contract No. 2007CB815004
and the Knowledge Innovation Project of Chinese Academy of Sciences
under Grant No. KJCX3.SYW.N2.


\end{document}